# Timing acquisition and demodulation of an UWB system based on the differential scheme

Karima Ben Hamida El Abri and Ammar Bouallegue
Syscoms Laboratory, National Engineering School of Tunis, Tunisia
Emails : enitkarima@yahoo.fr, ammar.bouallegue@enit.rnu.tn

## Abstract

Blind synchronization constitutes a major challenge in realizing highly efficient ultra wide band (UWB) systems because of the short pulse duration which requires a fast synchronization algorithm to accomodate several asynchronous users. In this paper, we present a new Code Block Synchronization Algorithm (CBSA) based on a particular code design for a non coherent transmission. Synchronization algorithm is applied directly on received signal to estimate timing offset, without needing any training sequence. Different users can share the available bandwith by means of different spreading codes with different lengths. This allows the receiver to separate users, and to recover the timing information of the transmitted symbols. Simulation results and comparisons validate the promising performance of the proposed scheme even in a multi user scenario. In fact, the proposed algorithm offers a gain of about 3 dB in comparaison with reference [5].

## 1   Introduction

In recent years, UWB technology has received great interests from industry as well as academia due to its many potential advantages, such as offering simultaneously high data rate on short distance, high resolution, low probability of detection and low power consumption. In order to meet the spectrum mask released by FCC and obtain adequate signal energy for reliable detection, each information symbol is represented by a train of very short pulses, called monocycles. Each located in its own frame. It was shown in [6], [7] and [11] that efficient demodulation of any UWB systems requires at the receiver an accurate timing estimation. This is due to the fact that the information bearing pulses are ultra short. Moreover, the dense multipath channel, through which these pulses propagate, is unknown at the receiver during the synchronization process. These reasons explain why synchronization has received so much importance in UWB research e.g. [3], [4], [5] and references therein. In this work, we consider blind synchronization since non data aided algorithms are desired in many potential UWB radio applications such as Wireless sensor and ad hoc networks where training sequence may not be available. [2] proposed a synchronization solution (TDT) that consists on the autocorrelation of consecutive symbol-long segments of the received UWB signal. It requires relatively long data records to reliably estimate the statistics on which it relies. In [3], a blind synchronization system (CABS) was proposed. It consists on changing the polarity of the pulses using carrefully designed codes that will be exploited to perform acquisition by correlating the received signal with the code template. It was shown that [3] outperforms [2]. Moreover, [2] and [3] both need long data records to achieve good performance which increases the complexity of the system. Capitalizing on the unique differential structure and the prior knowledge of the time hopping (TH) codes, [9] can achieve synchronization but it suffers from remarkable performance degradation because of the severe noise cross noise effect. In [5], the author propose an algorithm that relies only on the knowledge of the DS (direct sequence) codes and the signal structure. Timing acquisition is achieved via peak picking the objective function which is established over one symbol-long observation interval. It was shown in [5] that the algorithm proposed outperforms the algorithm in [9]. Thus, we compare our proposed approach to the method in [5]. In this paper, we propose a blind timing synchronization algorithm based on a particular code design of codes used in each symbol information.

As for demodulation, many receiver structure are proposed. The first one used is rake receiver which causes high receiver complexity because of the large number of fingers needed for the estimation [16]. As a solution, a transmitted reference (TR) receiver has been proposed [15]. In this case, the transmitted signal consists of a train of pulses pairs. Over each frame, the first pulse is modulated by data. The second one is a reference pulse used for signal detection at the receiver. Reception is made by delaying the received signal and correlating it with the original version. The simplicity of this receiver is very attractive. Nevertheless, TR systems waste half of the energy to transmit reference signals.

To overcome this problem, a differential system is proposed, where detection is achieved by correlating the received signal and its replica delayed by a period D (D can be the symbol period [12], the frame period [13] or a function of chip, symbol and frame period [13], [14]).

In this work, we propose to use a block differential system, where D is equal to the symbol period multiplied by the number of bits in each block of data.

The remainder of this paper is organized as follows. Section 2 gives a general description of the UWB channel model and the system model used in this work. Section 3 presents a description of our proposed solution. Then, we explain, in section 4, the choice of codes used, theoretically. In section 5, we present the receiver structure. Then, in section 6, we present an extension to the multi user case, followed by simulation results in section 7, and conclusion in section 8.

## 2 Modeling Preliminaries

### 2.1 UWB channel Model

The basic conditions of UWB systems differ according to applications. It is based on the conventionnal Saleh & Valenzuela (S-V) channel model [17]. We distinguish two kind of propagation environments : outdoor and indoor propagation. The former is dominated by a direct path while the latter is made of a dense multipath. In this work, we consider the IEEE 802.15 UWB indoor channel [8], where multipath arrivals are grouped into two categories :cluster arrivals, and ray arrivals within each cluster.

In this paper, we consider channel model 1 (CM1), representative of line of sight (LOS)(0-4m) channel conditions.

The channel impulse response is given by :

$$h(t) = \sum_{l=0}^{L-1} \sum_{k=0}^{K-1} \alpha_{k,l} \delta(t - T_l - \tau_{k,l}) \qquad (1)$$

where :
- $\alpha_{k,l}$ denotes the multipath gain coefficient.
- $T_l$ is the $l-th$ cluster arrival time.
- $\tau_{k,l}$ represents the delay of $k-th$ multipath component inside the cluster $l$.
- $\delta(t)$ is the Dirac delta function.

The UWB channel given in $(eq.1)$ can be modeled as a tapped delay line defined as follows :

$$h(t) = \sum_{l=0}^{L-1} \alpha_l \delta(t - \tau_l) \qquad (2)$$

with :
- $\alpha_l$ denotes attenuation of each path.
- $\tau_l$ represents the delay of $k-th$ path. It satisfies $\tau_0 < \tau_1 < \ldots < \tau_L$.

## 2.2 System Model

In the UWB transmission, every symbol is transmitted by employing $N_f$ short pulses $\omega_T(t)$, each with ultra short duration, $T_\omega$, of the order of nanosecond and normalized energy. The pulses are transmitted once per frame.
We propose to use a direct sequence DS-UWB system equipped with binary antipodal pulse amplitude modulation (PAM) that consists on multiplying pulses by a spreading sequence [1]. The transmitted signal is given by :

$$s(t) = \sum_m b_m \sum_{j=0}^{N_f - 1} C_{m,j} \omega_T(t - jT_f - mT_s) \qquad (3)$$

Where $\{b_m\}$ represents the random binary data symbol sequence taking values $\pm 1$, with equal probability. $T_f$ is the frame duration verifying $T_f >> T_\omega$ and $T_s$ is the symbol duration composed of $N_f$ frames : $T_s = N_f T_f$. $\{C_m\}$ is the code sequence assigned to the $m^{th}$ symbol.

Let's set the following definition :
$\omega_R(t) = \sum_{l=0}^{L} \alpha_l \omega_T(t - \tau_{l,0})$ which represents the received waveform without considering the timing offset of the first path $\tau_0$ ($\tau_{l,0} = \tau_l - \tau_0$).

The received signal can be written as :

$$r(t) = \sum_m b_m \omega_m(t - mT_s - \tau_0) + n(t) \qquad (4)$$

where $\omega_m(t) = \sum_{j=0}^{N_f - 1} C_{m,j} \omega_R(t - jT_f)$ represents the $m^{th}$ transmitted symbol waveform without modulation.

As for $n(t)$, it is assumed to be AWGN with power density $N_0/2$.

## 3 Description of the proposed synchronization algorithm

In this section, we introduce the idea of our algorithm which is a novel blind synchronization method to estimate the timing offset.

To encode data, we don't use a random sequence but a carrefully designed one. We choose the Hadamard code. So, the code assigned to the $i^{th}$ symbol $C_i$ is a Hadamard code of length $N_f$ taking values $\pm 1$. These codes are given by the rows of the Hadamard matrix $H$ of dimension $N_f^2$. Each row of H is orthogonal to all other rows. We propose to group information data in blocks containing M symbols each one. Consequently, we'll have $N/M$ blocks for $N$ transmitted bits (with N multiple of M). We use the same family of M orthogonal codes $\{C_i\}$ to spread the $M$ symbols contained in each block as presented in $Fig.1$.

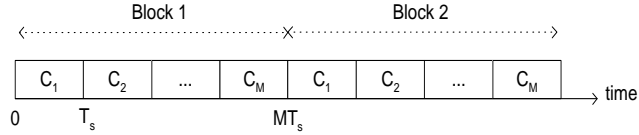

FIG. 1 – Example of two blocks of symbols

We propose to use a differential UWB system based on a differential transmission and detection.

So, the transmitted signal can be rewritten as follows :

$$s(t) = \sum_m d_m \sum_{j=0}^{N_f-1} C_{\lfloor \frac{m}{M} \rfloor, j} \omega_T(t - jT_f - mT_s) \tag{5}$$

Where :
- $\lfloor \cdot \rfloor$ is the modulus after division.
- $d_m$ is the block differentially encoded bit given by : $d_m = b_m . d_{m-M}$

In this case, the $m^{th}$ transmitted waveform becomes :
$\omega_m(t) = \sum_{j=0}^{N_f-1} C_{\lfloor \frac{m}{M} \rfloor, j} \omega_R(t - jT_f)$

Instead of taking M sequences, the transmitted signal is encoded using only (M-1) code sequences (M is the number of bits in each block).

In fact, one of these orthogonal sequences will be used to code two symbols in the same block to enable synchronization as shown in $Fig.2$ (codes $C_0$ and $C_u$ are identical). Then, in the other blocks, we use the same code's family.

To avoid correlation between codes in adjacent blocks and to enable multi users transmission, code distribution supposes a condition on the number of bits in each blocks. The criteria used in the choice of the appropriate family sequences will be mentionned latter.

Due to this code design and to the fact that codes are orthogonals, the transmitted symbol waveform obtained without modulation, $\omega_m(t)$ leads to the following properties :

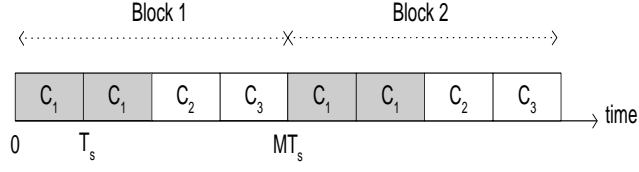

FIG. 2 – Example of sequences design for 2 blocks of symbols, M=4.

– Property 1 : Since codes are orthogonal :

$$\int_0^{T_s} \omega_m(t)\omega_{m'}(t)dt = N_f E_\omega \ \ if \ m = 0 \ and \ m' = m - 1$$
$$= 0 \ otherwise \qquad (6)$$

Where $E_\omega$ is the energy $\omega_m(t)$.

– Property 2 : The cross correlation between two given waveforms $\omega_m(t)$ and $\omega_{m'}(t - \tau')$ is non zero if property 1 is true and $|\tau'| \in [0, T_s)$.

In real UWB settings, the receiver knows neither the propagation delay $\tau_0$, nor the transmission starting time.
We suppose that the receiver initiates timing at $t_0$ ($t_0 \geq \tau_0$) . And, since $\tau_0$ serves only as a reference, we can set $\tau_0 = 0$. Hence, the received signal which initiates acquisition is given by :

$$\begin{aligned} y(t) &= r(t + t_0) \\ &= \sum_m b_m \omega_m(t - mT_s + t_0) + n(t + t_0) \end{aligned} \qquad (7)$$

To achieve synchronization, we correlate the received signal $y(t)$ with its replica delayed by $T_s$.

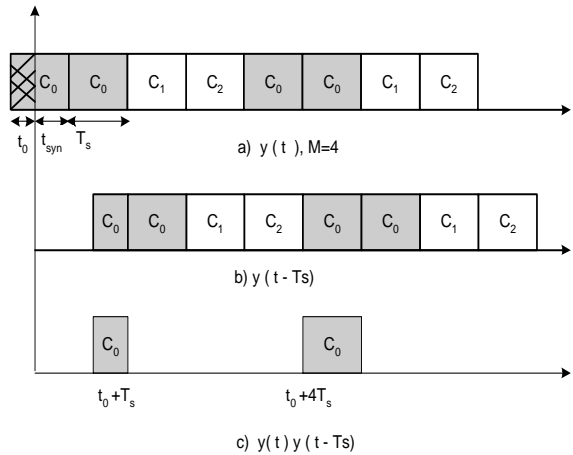

FIG. 3 – Illustration of the non-coherent synchronization

In $Fig.3$, we illustrate an example of the proposed structure. The origin of time is $t_0$, time when the receiver initiates acquisition. (a) represents the received signal $r(t)$, (b) is (a)'s $T_s$ delayed copy, and (c) is the product of (a) and (b). Since used codes are orthogonal, $<Ci, Cj>=\delta_{ij}$, resulting signal (c) is always null only in the regions $[t_0 + (n.M+1)T_s, t_0 + (n.M+2)T_s]$ for the $n^{th}$ block. Consequently, timing offset estimation is achieved when the maximum of energy in product signal is detected as we said above.

Our goal is to find the starting time of the next symbol coming after $t_0$, and therefore $\hat{t}_{sync}$. Then, we can simply find $\hat{t_0}$ :
$\hat{t}_0 = T_s - \hat{t}_{sync}$.

Assume that we process the received signal over a time interval of duration $2MT_s$, we estimate $t_{sync}$ by adjusting the observation window until reaching the maximum of the cross-correlation (see $Fig.3$).

Therefore, we can give the following criteria for our algorithm :

$$\hat{t}_{sync} = argmax_{\tau \in [0,T_s[} |J(\tau)| \qquad (8)$$

where :

$$\begin{aligned} J(\tau) &= \int_{\tau}^{\tau+T_s} y(t)y(t-T_s)dt \\ &= E_\omega \{-D_{k+1} \sum_{j=0}^{j_0} C_{\lfloor \frac{k+1}{M} \rfloor,j} C_{\lfloor \frac{k}{M} \rfloor,j} + D_{k+2} \sum_{j=0}^{j_0} C_{\lfloor \frac{k+2}{M} \rfloor,j} C_{\lfloor \frac{k+1}{M} \rfloor,j} \} \\ &+ \{D_{k+1} C_{\lfloor \frac{k+1}{M} \rfloor,j_0} C_{\lfloor \frac{k}{M} \rfloor,j_0} - D_{k+2} C_{\lfloor \frac{k+2}{M} \rfloor,j_0} C_{\lfloor \frac{k+1}{M} \rfloor,j_0} \} \int_{\delta}^{T_f} \omega_R^2(t)dt \end{aligned} \qquad (9)$$

(see $Appendix$)

## 4 Code design

### 4.1 Choice of sequences

The predominant term in the expression of $J(\tau)$ in equation 9 are parts one and two since $\int_{\delta}^{T_f} \omega_R^2(t)dt$ is zero unless if : $0 \leq \delta < T^{corr}$.

Consequently, to maximize $J(\tau)$, we have to minimize the following term :

$$J'(\tau) = -D_{k+1} \sum_{j=0}^{j_0} C_{\lfloor \frac{k+1}{M} \rfloor,j} C_{\lfloor \frac{k}{M} \rfloor,j} + D_{k+2} \sum_{j=0}^{j_0} C_{\lfloor \frac{k+2}{M} \rfloor,j} C_{\lfloor \frac{k+1}{M} \rfloor,j} \qquad (10)$$

$$|J'(\tau)| \leq \left| \sum_{j=0}^{j_0} C_{\lfloor \frac{k+1}{M} \rfloor,j} C_{\lfloor \frac{k}{M} \rfloor,j} \right| + \left| \sum_{j=0}^{j_0} C_{\lfloor \frac{k+2}{M} \rfloor,j} C_{\lfloor \frac{k+1}{M} \rfloor,j} \right| \qquad (11)$$

Our goal is to minimize this upper bound. To this end, we have to choose the appropriate sequences to encode symbols so that the correlation between codes $C_n$ and $C_{n'}$ is nearly equal to $0$. Thus, we specify the following criteria for a given code couple $(C_n, C_{n'})$ :

$$S_{n,n'} = Max_{j_0} \sum_{j=0}^{j_0} C_{n,j} C_{n',j} \qquad (12)$$

First, we calculate $S_{n,n'}$ for all existing code couple. Then, we select codes having the minimum $S_{n,n'}$.

## 4.2 Synchronization process

First, instead of considering only one block to estimate $t_{syn}$, we propose to sum the energies contained in $B$ blocks.
In this case, the expression of $J(\tau)$ is given by :

$$\begin{aligned}J(\tau) &= B\{E_\omega\{-D_{k+1}\sum_{j=0}^{j_0} C_{\lfloor\frac{k+1}{M}\rfloor,j}C_{\lfloor\frac{k}{M}\rfloor,j} + D_{k+2}\sum_{j=0}^{j_0} C_{\lfloor\frac{k+2}{M}\rfloor,j}C_{\lfloor\frac{k+1}{M}\rfloor,j}\} \\ &+ \{D_{k+1}C_{\lfloor\frac{k+1}{M}\rfloor,j_0}C_{\lfloor\frac{k}{M}\rfloor,j_0} - D_{k+2}C_{\lfloor\frac{k+2}{M}\rfloor,j_0}C_{\lfloor\frac{k+1}{M}\rfloor,j_0}\}\int_\delta^{T_f}\omega_R^2(t)dt\}\end{aligned} \quad (13)$$

Synchronization process includes two stages.
In stage one, we aim to obtain a coarse estimation of $t_{sync}$. The search is done using the step $T_e = T_\omega$ from the begining of the received signal $r(t)$. For each value, we calculate $J(\tau)$ during an integration window of length $T_s$. Then, we select the maximum of all the integrators which corresponds to $J(\tau_1)$.
Once $\tau_1$ found, we move to the second stage in which we perform a smooth search by a step of $T_e = 1$. Like we have done in stage 1, we calculate $J(\tau)$ during an integration window of length $T_s$ starting from $\tau 1$ to $T_s$. Then, we select the time which corresponds to $J(\hat{t}_{sync})$.
Finally, we juste have to make difference between $T_s$ and $\hat{t}_{sync}$ to find the estimation of the starting time of the receiver $\hat{t}_0$.

## 5 Non coherent differential detection

Once synchronization established, the received signal at the receiver input is given by :

$$\begin{aligned}X(t) &= y(t - \hat{t}_0) \\ &= \sum_m d_m\omega_m(t - mT_s + t_0 - \hat{t}_0) + n(t + t_0 - \hat{t}_0)\end{aligned} \quad (14)$$

We suppose perfect timing. In this case, we have :

$$\begin{aligned}X(t) &= \sum_m d_m\omega_m(t - mT_s) + n(t) \\ &= s(t) + n(t)\end{aligned} \quad (15)$$

With this notation $s(t)$ is the desired signal and $n(t)$ represents noise.
To detect the emitted symbols $b_m$, we suggest to use a differential receiver based on the correlation of the received signal, given in equation 15,with its replica delayed by a block ($MT_s$) since data are coded differentially by block.
A block diagram of the differential receiver is presented in $Fig.4$.

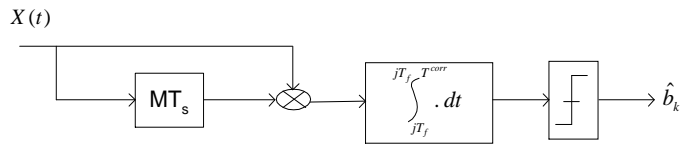

FIG. 4 – Differential receiver structure

The bloc differential receiver correlates the received signal with its replica delayed by $MT_s$. The output of the correlator for the $k^{th}$ symbol is given by the expression below :

$$\begin{aligned} X_k &= \int_{kT_s}^{(k+l)T_s} X(t)X(t-MT_s)dt \\ &= s(k) + \sum_{i=1}^{3} n_i(k) \end{aligned} \quad (16)$$

Where $\{n_i(k)\}_{i=1\ldots 3}$ are noise terms due to the correlation between desired signal and noise. Table (1) summarizes these terms.

| $\int_{kT_s}^{(k+1)T_s} dt$ | s(t) | n(t) |
|---|---|---|
| $s(t-MT_s)$ | s(k) | $n_2(k)$ |
| $n(t-MT_s)$ | $n_1(k)$ | $n_3(k)$ |

TAB. 1 – Signal and noise terms at the correlator output.

The desired signal is given by :

$$\begin{aligned} s(k) &= \int_{kT_s}^{(k+l)T_s} s(t)s(t-MT_s)dt \\ &= \sum_{m,m'} d_m d_{m'} \int_{kT_s}^{(k+l)T_s} \omega_m(t-mT_s)\omega_{m'}(t-(m'+M)T_s)dt \\ &= \sum_m d_m d_{m-M} \int_{kT_s}^{(k+l)T_s} \omega_m(t-mT_s)\omega_{m-M}(t-mT_s)dt \\ &= \sum_m b_m \int_{kT_s}^{(k+l)T_s} \omega_m(t-mT_s)\omega_{m-M}(t-mT_s)dt \\ &= \sum_m b_m \int_0^{T_s} \omega_m(t+(k-m)T_s)\omega_{m-M}(t+(k-m)T_s)dt \\ &= b_k \int_0^{T_s} \omega_k(t)\omega_{k-M}(t)dt \\ &= b_k \sum_{j=0}^{N_f-1} \int_{jT_f}^{jT_f+T^{corr}} \sum_{j',j''} C_{\lfloor \frac{k}{M} \rfloor, j'} C_{\lfloor \frac{k-M}{M} \rfloor, j''} \omega_R(t-j'T_f)\omega_R(t-j''T_f)dt \\ &= b_k \sum_{j=0}^{N_f-1} C_{\lfloor \frac{k}{M} \rfloor, j} C_{\lfloor \frac{k-M}{M} \rfloor, j} \int_{jT_f}^{jT_f+T^{corr}} \omega_R^2(t-jT_f)dt \\ &= N_f b_k E_R \end{aligned} \quad (17)$$

Where :
- $T^{corr}$ is the integration time of the correlator. It is choosen to be $T^{corr} = T_\omega + T^{mds}$ ($T^{mds}$ is the maximum delay spread of the channel).
- $E_R$ is the energy contained in the received waveform $\omega_R(t)$.

To decide of the value of the $k^{th}$ transmitted bit, we just have to see the sign of $X_k$.
$\hat{b}_k = sign(X_k)$

## 6 Extension to the multiuser case

We want to use the same transmission scheme for multiple access and the same receiver structure.
First, the number of bits in each symbol must verify the following property : $M > 2U$ to enable multi user transmission.

We use $(M-1)$ different code sequences for each user. The location of each code is described in $Fig.5$ (codes $C_1$ and $C_u$ are identical).

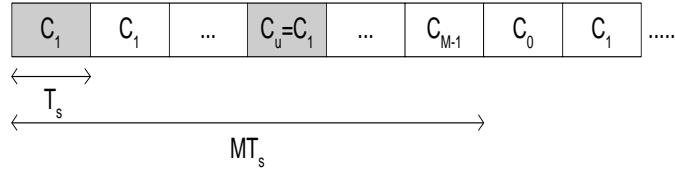

FIG. 5 – Example of sequences design.

Then, to synchronize the $u^{th}$ user's signal, we just have to correlate the received signal with its replica delayed by $uT_s$.
Another problem can overcome from the fact that when multiple users are active in the channel, their pulses can be supperposed. In our work, template waveform used in synchronization phase is the delayed replica of the received signal. Therefore, all users signal are detected at the same time.
On the other hand, we want to discriminate different active users. And on the other hand, we want to keep a simple synchronization structure.
And so, we have a compromise between complexity and performance to satisfy.
We propose to fix the symbol period $T_s$ for all users, and to define a different frame number for each user. Consequently, pulse repetition period is variable and verifies the following condition :
$T_s = N_f^1 T_f^1 = N_f^2 T_f^2 = ... = N_f^U T_f^U$
Where U is the number of active users in tha channel.

Simulation results of the proposed system are presented in the next section.

# 7 Simulation Results and Comparison

This section is devoted to the presentation of some simulations to test the performance of the proposed algorithm in terms of probability of acquisition, normalized mean squre error (MSE) and detection capability.
We also make a comparaison with the solution presented in [5] through numerical Monte Carlo simulations.

## 7.1 Single user system

In each trial, the following suppositions are made :

- The monocycle $\omega_T(t)$ is chosen as the second derivate of a Gaussian pulse, with unit energy and duration $T_\omega = 1ns$.
- The number of frame in each symbol $N_f = 16$ frames, and the duration of each one is $T_f = 10ns$.

- Spreading codes are choosen so that they verify the criteria in eq. 12. In this simulations, we used the hadamard codes of length $N_f$.
  The DS codes in [5] are generated randomly from $\pm 1$, with equal probability.
- The number of bits in each block M=5.
- The timing offset $t_0$ is randomly generated from a uniform distribution over $[0, T_s)$.
- We simulate the multipath channel using the model CM1 from [8].
  The channel is assumed to be time invariant within a burst of symbols.
  The maximum delay spread of the channel is $10ns$

- The number of bits over which we decide is $B = 4$.

We begin by evaluating the performance in terms of probability of acquisition as a function of the signal-to-noise (SNR) ratio, which is the energy per symbol over the noise power $N_0$. The results are given in $Fig.6$.

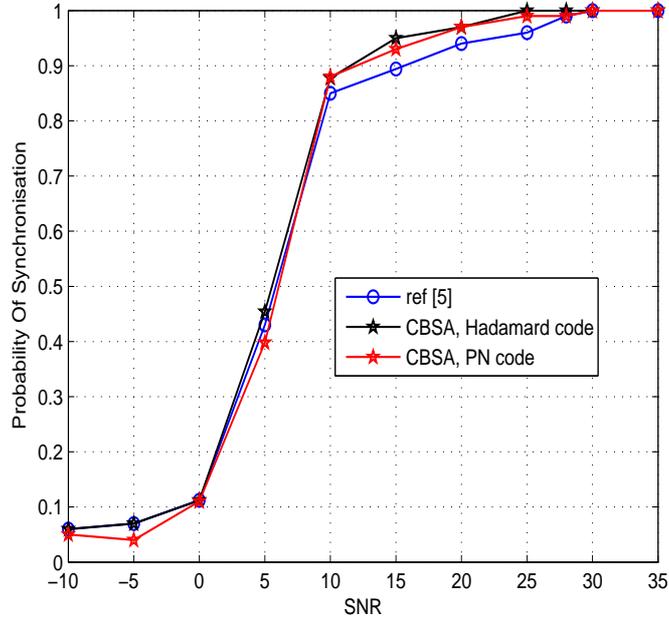

FIG. 6 – Probability of acquisition, B=4

We can see that both algorithms can achieve synchronization but the acquisition probability of our algorithm (BCSA) outperforms [5] in acquisition and offers a gain of about 3 dB.
We have also testing the robustness of the BCSA algorithm against codes used, which are Hadamard codes and Pseudo Noise codes. We constate that the two types of codes used offers nearly the same results.

Then, we carry out Normalized MSE comparison which is given by $E\left\{(\hat{\tau} - \tau)\right\}^2$ normalized to $T_s^2$.
$Fig.7$ draws the results of the two techniques. As SNR increases, the NMSE decreases and curves converges since $\hat{t_0}$ converges to $t_0$.
We can also see that our algorithm outperforms [5]. Through simulations performed, we conclude that our solution offers a gain of about 3 dB in comparison with [5]. In fact, [5] needs large number of bits to achieve acquisition.

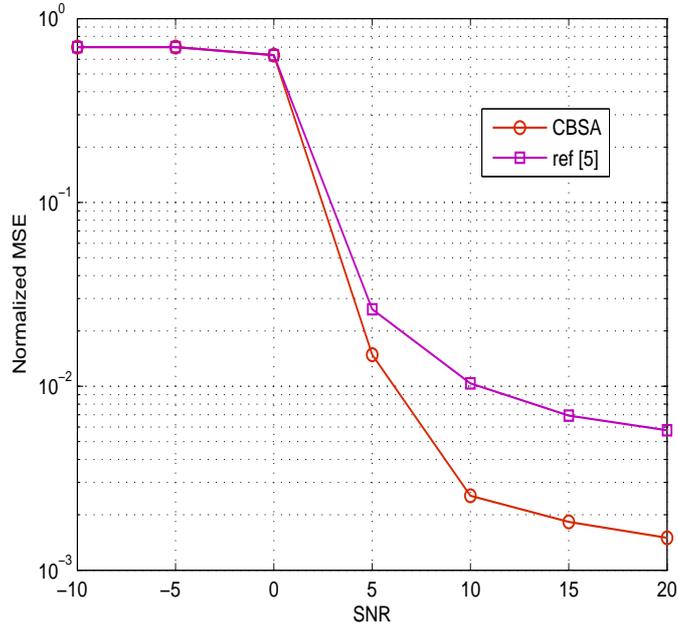

FIG. 7 – Normalized MSE

## 7.2 Multi user system

In this paragraph, we investigate the performance of our system to the multi user case. Parameters used are as follows :

- The number of frame $N_f^u$ in each user's symbol are 32, 21 and 15, respectively.
- The number of users in the system U=3.
- The number of bits in each block M=7 ($M \geq 2U$).
- We used the orthogonal PN codes of length $N_f$.
  The DS codes in [5] are generated randomly from $\pm 1$, with equal probability.
- The number of bits over which we decide is $B = 4$.

Results of the simulations are drawn in $Fig.8$ and $9$.
We can affirm through simulations that the system proposed is robust to multi user interference.

## 8 Conclusion

In this paper, a new solution called $BCSA$ is proposed for blind synchronization for a DS-UWB system. It is based on a judicious design of the spreading sequence used in each block of symbols. Simulation results show a good performance of our method that outperforms algorithm described in [5] and provides a gain of 3 dB. The system proposed is valid even in a multiple access system.

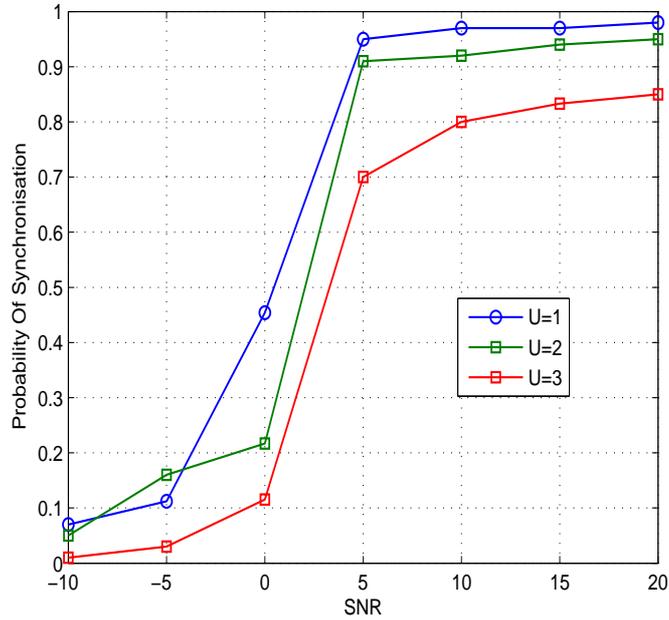

FIG. 8 – Probability of acquisition, B=4

# Appendix

### Proof of equation [9]

In this section, we simplify the expression of $J(\tau)$. For brievity, we consider only the useful part of the product of $y(t)$ and $y(t - T_s)$ denoted $x(t)$. First, we will simplify the expression of $x(t)$.

$$x(t) = \sum_{m,m'} b_m b_{m'} \omega_m(t - mT_s + t_0)\omega_{m'}(t - (m'+1)T_s + t_0) \quad (18)$$

Using property 2, the integration of $x(t)$ will be zero unless if :
$0 \leq |mT_s - t_0 - (m'+1)T_s + t_0| < T_s$
Consequently, $m' = m - 1$
Using this last expression and (18), (9) can be rewritten as :

$$\begin{aligned} J(\tau) &= \int_{\tau}^{\tau+T_s} x(t)dt \\ &= \sum_m D_m \int_{\tau}^{\tau+T_s} \gamma_m(t - mT_s + t_0)dt \end{aligned} \quad (19)$$

Where :
- $D_m = b_m b_{m-1}$
- $\gamma_m(t) = \omega_m(t)\omega_{m-1}(t)$

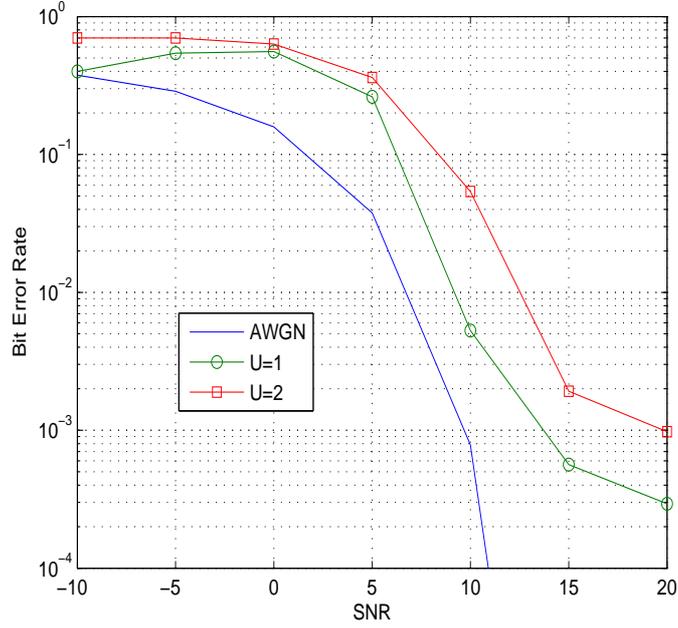

FIG. 9 – Bit Error Rate

Let's pose :

$$\begin{aligned} I_m &= \int_{\tau}^{\tau+T_s} \gamma_m(t - mT_s + t_0)dt \\ &= \int_{\tau}^{\tau+T_s} \gamma_m(t - (m-1)T_s - t_{sync})dt \\ &= \int_{\tau-t_{sync}}^{\tau-t_{sync}+T_s} \gamma_m(t - (m-1)T_s)dt \end{aligned} \quad (20)$$

$\tau - t_{sync}$ can be developped as follows :

$$\tau - t_{sync} = kT_s + \beta \quad (21)$$

Where $k$ and $\beta$ represent the time delay at symbol level and frame level, respectively ($\beta \in [0, T_s[$). As a result :

$$\begin{aligned} I_m &= \int_{kT_s+\beta}^{(k+1)T_s} \gamma_m(t-(m-1)T_s)dt + \int_{(k+1)T_s}^{(k+1)T_s+\beta} \gamma_m(t-(m-1)T_s)dt \\ &= \int_{\beta}^{T_s} \gamma_m(t+(k-m+1)T_s)dt + \int_0^{\beta} \gamma_m(t+(k-m+2)T_s)dt \end{aligned} \quad (22)$$

Consequently, we have :

$$\begin{aligned} J(\tau) &= D_{k+1} \int_{\beta}^{T_s} \gamma_{k+1}(t)dt + D_{k+2} \int_0^{\beta} \gamma_{k+2}(t)dt \\ &= D_{k+1}V_k(\beta) + D_{k+2}W_k(\beta) \end{aligned} \quad (23)$$

Where $V_k(\beta)$ and $W_k(\beta)$ are developped above :

$$\begin{aligned} V_k(\beta) &= \int_{\beta}^{T_s} \gamma_{k+1}(t)dt \\ &= \sum_{j=0}^{N_f-1} C_{\lfloor \frac{k+1}{M} \rfloor, j} C_{\lfloor \frac{k}{M} \rfloor, j} \int_{\beta}^{T_s} \omega_R^2(t - jT_f)dt \end{aligned} \quad (24)$$

Let's pose : $\beta = j_0 T_f + \delta$, where $\delta = \beta - j_0 T_f \in [0, T_f[$.
The last one represents the time delay at pulse level.

Consequently, (24) becomes :

$$V_k(\beta) = \sum_j C_{\lfloor \frac{k+1}{M} \rfloor,j} C_{\lfloor \frac{k}{M} \rfloor,j} \{ \int_\beta^{(j_0+1)T_f} \omega_R^2(t-jT_f)dt + \int_{(j_0+1)T_f}^{T_s} \omega_R^2(t-jT_f)dt \} \quad (25)$$

The second integral in (25) is non zero only if :
$j_0 + 1 \leq j < N_f$
In this case, (25) can be rewritten as follows :

$$\begin{aligned} V_k(\beta) &= \sum_{j=0}^{N_f-1} C_{\lfloor \frac{k+1}{M} \rfloor,j} C_{\lfloor \frac{k}{M} \rfloor,j} \\ &\int_{j_0 T_f+\delta}^{(j_0+1)T_f} \omega_R^2(t-jT_f)dt + \sum_{j=j_0+1}^{N_f-1} C_{\lfloor \frac{k+1}{M} \rfloor,j} C_{\lfloor \frac{k}{M} \rfloor,j} \int_{(j_0+1)T_f}^{T_s} \omega_R^2(t-jT_f)dt \\ &= C_{\lfloor \frac{k+1}{M} \rfloor,j_0} C_{\lfloor \frac{k}{M} \rfloor,j_0} \int_\delta^{T_f} \omega_R^2(t)dt + E_\omega \sum_{j=j_0+1}^{N_f-1} C_{\lfloor \frac{k+1}{M} \rfloor,j} C_{\lfloor \frac{k}{M} \rfloor,j} \end{aligned} \quad (26)$$

As for $W_k(\beta)$, it's given by :

$$\begin{aligned} W_k(\beta) &= \int_0^\beta \gamma_{k+2}(t)dt \\ &= \sum_{j=0}^{N_f-1} C_{\lfloor \frac{k+2}{M} \rfloor,j} C_{\lfloor \frac{k+1}{M} \rfloor,j} \int_0^\beta \omega_R(t-jT_f)dt \end{aligned} \quad (27)$$

With $\beta = (j_0+1)T_f - \delta'$, $\delta' \in [0, T_f[$.
So, (27) can be rewritten as :

$$W_k(\beta) = \sum_{j=0}^{N_f-1} C_{\lfloor \frac{k+2}{M} \rfloor,j} C_{\lfloor \frac{k+1}{M} \rfloor,j} \{ \int_0^{(j_0+1)T_f} \omega_R^2(t-jT_f)dt - \int_{(j_0+1)T_f-\delta'}^{(j_0+1)T_f} \omega_R^2(t-jT_f)dt \} \quad (28)$$

The primal integral in (28) is non zero only if :
$(j_0+1)T_f \geq jT_f + T^{corr} \Leftrightarrow j \leq (j_0+1) - \frac{T^{corr}}{T_f} \Rightarrow j \leq j_0$
Where $T^{corr}$ is the integration time of the correlator. It's choosen to be $T^{corr} = T_\omega + T_{mds}$ ($T_{mds}$ is the maximum delay spread of the channel). And when $T^{corr}$ is unknown, it can be replaced by an upper bound or even $T_f$.
Consequently, (28) becomes :

$$\begin{aligned} W_k(\beta) &= \sum_{j=0}^{j_0} C_{\lfloor \frac{k+2}{M} \rfloor,j} C_{\lfloor \frac{k+1}{M} \rfloor,j} \int_0^{(j_0+1)T_f} \omega_R^2(t-jT_f)dt \\ &- \sum_{j=0}^{N_f-1} C_{\lfloor \frac{k+2}{M} \rfloor,j} C_{\lfloor \frac{k+1}{M} \rfloor,j} \int_{(j_0+1)T_f-\delta'}^{(j_0+1)T_f} \omega_R^2(t-jT_f)dt \\ &= E_\omega \sum_{j=0}^{j_0} C_{\lfloor \frac{k+2}{M} \rfloor,j} C_{\lfloor \frac{k+1}{M} \rfloor,j} - C_{\lfloor \frac{k+2}{M} \rfloor,j_0} C_{\lfloor \frac{k+1}{M} \rfloor,j_0} \int_{T_f-\delta'}^{T_f} \omega_R^2(t)dt \\ &= E_\omega \sum_{j=0}^{j_0} C_{\lfloor \frac{k+2}{M} \rfloor,j} C_{\lfloor \frac{k+1}{M} \rfloor,j} - C_{\lfloor \frac{k+2}{M} \rfloor,j_0} C_{\lfloor \frac{k+1}{M} \rfloor,j_0} \int_\delta^{T_f} \omega_R^2(t)dt \end{aligned} \quad (29)$$

Replacing (26) and (29) in (23), we obtain :

$$\begin{aligned} J(\tau) &= D_{k+1} \{ C_{\lfloor \frac{k+1}{M} \rfloor,j_0} C_{\lfloor \frac{k}{M} \rfloor,j_0} \int_\delta^{T_f} \omega_R^2(t)dt + E_\omega \sum_{j=j_0+1}^{N_f-1} C_{\lfloor \frac{k+1}{M} \rfloor,j} C_{\lfloor \frac{k}{M} \rfloor,j} \} \\ &+ D_{k+2} \{ E_\omega \sum_{j=0}^{j_0} C_{\lfloor \frac{k+2}{M} \rfloor,j} C_{\lfloor \frac{k+1}{M} \rfloor,j} - C_{\lfloor \frac{k+2}{M} \rfloor,j_0} C_{\lfloor \frac{k+1}{M} \rfloor,j_0} \int_\delta^{T_f} \omega_R^2(t)dt \} \\ &= E_\omega \{ D_{k+1} \sum_{j=j_0+1}^{N_f-1} C_{\lfloor \frac{k+1}{M} \rfloor,j} C_{\lfloor \frac{k}{M} \rfloor,j} + D_{k+2} \sum_{j=0}^{j_0} C_{\lfloor \frac{k+2}{M} \rfloor,j} C_{\lfloor \frac{k+1}{M} \rfloor,j} \} \\ &+ \{ D_{k+1} C_{\lfloor \frac{k+1}{M} \rfloor,j_0} C_{\lfloor \frac{k}{M} \rfloor,j_0} - D_{k+2} C_{\lfloor \frac{k+2}{M} \rfloor,j_0} C_{\lfloor \frac{k+1}{M} \rfloor,j_0} \} \int_\delta^{T_f} \omega_R^2(t)dt \end{aligned} \quad (30)$$

Since the $(M-1)$ codes used are orthogonal, we can ascertain the following egality :

$$\sum_{j=0}^{j_0} C_{\lfloor \frac{k+1}{M} \rfloor,j} C_{\lfloor \frac{k}{M} \rfloor,j} = - \sum_{j=j_0+1}^{N_f-1} C_{\lfloor \frac{k+1}{M} \rfloor,j} C_{\lfloor \frac{k}{M} \rfloor,j} \qquad (31)$$

Using (31) in (30), we obtain :

$$\begin{aligned} J(\tau) &= E_\omega \{ -D_{k+1} \sum_{j=0}^{j_0} C_{\lfloor \frac{k+1}{M} \rfloor,j} C_{\lfloor \frac{k}{M} \rfloor,j} + D_{k+2} \sum_{j=0}^{j_0} C_{\lfloor \frac{k+2}{M} \rfloor,j} C_{\lfloor \frac{k+1}{M} \rfloor,j} \} \\ &+ \{ D_{k+1} C_{\lfloor \frac{k+1}{M} \rfloor,j_0} C_{\lfloor \frac{k}{M} \rfloor,j_0} - D_{k+2} C_{\lfloor \frac{k+2}{M} \rfloor,j_0} C_{\lfloor \frac{k+1}{M} \rfloor,j_0} \} \int_\delta^{T_f} \omega_R^2(t) dt \end{aligned} \qquad (32)$$

# Références

[1] N. Boubaker and K. B. Letaief, "Ultra Wideband DSSS for Multiple Access Communications Using Antipodal Signaling", Communications, 2003. ICC 03. IEEE International Conference on, Volume : 3, On page(s) : 2197- 2201 vol.3.

[2] L.Yang and G. B. Giannakis, "Timing Ultra-Wideband Signals With Dirty Templates", in Proc. IEEE Trans. Commun., VOL.53, no.11, NOVEMBER 2005.

[3] M. Ghogho and Y. Ying,"Code-Assisted Blind Synchronization for UWB Systems", in Proc. IEEE ICC 06, Istambul, Turkey, June 2006, pp. 5080-5085.

[4] Y. Qiao, T. Lv, and L. Zhang, "A new blind synchronization algorithm for UWB-IR systems", in proc. IEEE VTCŠ09, Spain, April 2009, pp.1-5.

[5] Y. Qiao and T. Lv, "Blind Synchronization and low complexity Demodulation for DS-UWB systems", in proc. IEEE WCNC, 2010.

[6] Z. Tian and G. B. Giannakis, "BER sensitivity to mistiming in ultrawideband impulse radios- part I : Nonrandom channels", IEEE Trans. on Sig. Process., vol. 53, no. 4, pp. 1550-1560, April 2005.

[7] Z. Tian and G. B. Giannakis, "BER sensitivity to mistiming in ultrawideband impulse radios- part II : fading channels", IEEE Trans. on Sig. Process., vol. 53, no. 5, pp. 1897-1907, May 2005.

[8] J. R. Foerster and Q. Li, "Uwb channel modeling contribution from intel",IEEE 802.15.3 Wireles Personal Area Networks, Tech. Rep. IEEE p802.15-02/279r0-SG3a,Jun. 2002.

[9] A. A. D. Amico, U. Mengali, and L. Taponecco, "Synchronization for differential transmitted reference uwb receivers", IEEE Trans. Wireless Commun., vol. 6, no. 11, pp. 154-163, Nov. 2007.

[10] L. Yang and G.B. Giannakis, "Ultra wideband communications", IEEE Signal Processing Magazine, vol. 21, no. 6, pp. 26-54, 2004.

[11] G.F Tchev, P. Ubolkosold, S. Knedlik, O. Loffeld and K. Witrisal, "Data aided timing acquisition in uwb differential transmitted reference systems", PIMRC, 2006.

[12] Pausini M. and Janssen G.J.M. ; "Analysis and comparison of autocorrelation receivers for IR-UWB signals based on differential detection", Acoustics, Speech, and Signal Processing, 2004. Proceedings. (ICASSP 04). IEEE International Conference on, Volume 4, 17-21 May 2004 Page(s) :iv-513 - iv-516 vol.4

[13] Witrisal K. and Pausini M. ; "Equivalent system model of ISI in a framedifferential IR-UWB receiver", Global Telecommunications Conference, 2004. GLOBECOM Š04. IEEE, Volume 6, 29 Nov.-3 Dec. 2004 Page(s) :3505 - 3510 Vol.6

[14] Durisi G. and Benedetto S., "Performance of coherent and noncoherent receivers for UWB communications", Communications, 2004 IEEE International Conference on Volume 6, 20-24 June 2004 Page(s) :3429 - 3433 Vol.6

[15] Gezici S., Tufvesson F. and Molisch A.F., "On the performance of transmitted-reference impulse radio", Global Telecommunications Conference, 2004. GLOBECOM 04. IEEE Volume 5, 29 Nov.-3 Dec. 2004 Page(s) :2874 - 2879 Vol.5

[16] D. Cassioli, M. Z.Win, F. Vatalaro, and A. F. Molisch, "Performance of Low-Complexity Rake Reception in a Realistic UWB Channel", Proc. International Conference on Communications, New York, pp. 763-767, Apr. 28/May 2, 2002.

[17] Adel A. M. Saleh et REINALDO A. Valenzuela, "A Statistical Model for Indoor Multipath Propagation", IEEE Journal on selected Areas in COMMUNICATIONS. VOL. SAC-5. NO. 2. FEBRUARY 1987.

[18] S. Dasand and B. Das,"A compapision study of time domain equalization technique using UltraWide Band receivers performance for high data rate WPAN system", International Journal of Computer Networks & Communications (IJCNC), Vol.2, No.4, July 2010.